\documentclass[reprint,amsmath,amssymb,aps,prb,showpacs,floatfix,superscriptaddress]{revtex4-1}

\usepackage{graphicx}
\usepackage{bm}
\usepackage{hyperref}
\usepackage{verbatim}

\usepackage{color}          % for color
\usepackage[normalem]{ulem} % four sout{} command
\newcommand{\figureref}[1]{Fig.~\ref{#1}}
\renewcommand{\figurename}{Fig.}

\newcommand{\tr}{\mathrm{tr}}
\newcommand{\te}[1]{\mathrm{#1}}
\newcommand{\bra}[1]{\langle #1 |}
\newcommand{\ket}[1]{| #1 \rangle}

\newcommand{\f}[1]{#1}
\renewcommand{\Re}{\operatorname{Re}}

\newcommand\abs[1]{\lvert#1\rvert}

\newcommand\avgs[1]{\langle#1\rangle}
\newcommand{\dif}{\mathrm{d}}
\newcommand{\pd}{\partial}

\newcommand{\cd}{c^{\dag}}
\newcommand{\can}{c^{\phantom{\dag}}}
\newcommand{\dd}{d^{\dag}}
\newcommand{\dan}{d^{\phantom{\dag}}}

\newcommand{\peak}{\mathrm{peak}}

\newcommand{\subdot}{{}}
\newcommand{\subdotc}{{}}
\newcommand{\mc}{\mathcal}
\newcommand{\Real}{\operatorname{Re}}
\newcommand{\Imag}{\operatorname{Im}}
\begin{document}

\title{Violation of Onsager's theorem in approximate master equation approaches}

\author{Kevin Marc Seja}
\affiliation{Mathematical Physics and NanoLund, University of Lund, Box 118, 22100 Lund, Sweden}
\affiliation{Institute of Theoretical Physics, Technische Universit\"{a}t Dresden, 01062 Dresden, Germany}%

\author{Gediminas Kir{\v{s}}anskas}
\affiliation{Mathematical Physics and NanoLund, University of Lund, Box 118, 22100 Lund, Sweden}

\author{Carsten Timm}
\affiliation{Institute of Theoretical Physics, Technische Universit\"{a}t Dresden, 01062 Dresden, Germany}%

\author{Andreas Wacker}
\affiliation{Mathematical Physics and NanoLund, University of Lund, Box 118, 22100 Lund, Sweden}

\date{\today}

\begin{abstract}
The consistency with Onsager's theorem is examined for commonly used perturbative approaches, such as the Redfield and second-order von Neumann master equations, for thermoelectric transport through nanostructures. We study a double quantum dot, which requires coherences between states for a correct description, and we find that these perturbative approaches violate Onsager's theorem. We show that the deviations from the theorem scale with the lead-coupling strength in an order beyond the one considered systematically in the respective approach.
\end{abstract}

%\pacs{Valid PACS appear here}
\maketitle

\section{Introduction}

Understanding transport in nanoscale systems is crucial for applications ranging from nanoscale electronics to high-efficiency thermoelectric devices \cite{DresselhausAdvMater2007}. With experimental progress over the last decade, more and more theoretical suggestions become realizable in practice \cite{Martinez-BlancoNatPhys2015}. Moreover, transport in nanostructures provides an ideal stage for the study of the fundamental physics of open quantum systems far from equilibrium. For device design and optimization, as well as for the study of fundamental questions, a reliable theory for charge and energy transport in nanostructures is essential.

One such theory is the master-equation (ME) approach. Formally, MEs rely on a perturbative expansion in the coupling strength between a quantum system and its environment, e.g., connected leads. It has been shown for non-interacting systems that even such an approximate treatment can give charge currents that agree with an exact calculation \cite{KarlstromJPA2013, KonigPRB1996}. However, the predictions for energy transport have not been examined as much, although this is {also} important for applications.

The Wangsness-Bloch-Redfield (WBR) equation \cite{BlochPR1957,RedfieldIBM1957,RedfieldAdvMagnOptReson1965,WangsnessPR1953,TimmPRB2008} is a frequently used variant of the ME. It has long been known that, without a rotating-wave approximation, this equation is not of Lindblad form \cite{LindbladCMP1976} and does not conserve the positivity of the reduced density matrix \cite{BreuerBook2006}. In some cases this can lead to unphysical behavior such as large negative currents \cite{GoldozianSciRep2016}. A rotating-wave approximation corresponds to neglecting all off-diagonal elements \cite{BreuerBook2006}, which suggests that the behavior of the WBR equation is very much depending on the treatment of coherences. More recently, Hussein and Kohler \cite{HusseinPRB2014} found that a full WBR equation including coherences predicts charge currents that are not consistent with the exchange fluctuation theorems \cite{CampisiRMP2011}. In this paper, we examine the influence of coherences on the energy current predicted by such approaches. A special focus is on whether the currents satisfy certain Onsager relations \cite{OnsagerPR1931,OnsagerPR1931a}. As a direct consequence of Onsager's theorem from 1931, such relations between linear response functions have been a cornerstone of non-equilibrium thermodynamics for decades.
We solve the MEs numerically and, for the non-interacting limit, compare analytical results to the exact transmission formalism \cite{CaroliJPhysCSolidStatePhys1971,ButtikerPRB1985,DattaBook1995}.

The paper is organized as follows. In Section II the model for the spin-polarized two-level quantum dot and the considered Onsager relations are introduced.  Numerical results for the violation of Onsager's theorem using various approximate master equation approaches are presented in Section III. An analytical examination of the violation in the non-interacting case $U=0$ is given in Section IV and the scaling behavior of the violation with the coupling strength $\Gamma$ is discussed in Section V. Concluding remarks are given in Section VI. We present more explicit equations for the Redfield, first-order von Neumann (1vN), and Pauli master-equation approaches, and briefly discuss the second-order von Neumann approach (2vN) in Appendices A, B, and C. Furthermore, we provide a short derivation of the currents for the non-interacting case of $U=0$, using both the transmission-function formalism (Appendix D) and first-order master equations (Appendix E). Throughout the paper our units are such that $\hbar=e=k_{\mathrm{B}}=1$.

%%%%%%%%%%%%%%%%%%%%%%%%%%%%%%%%%%%%%%%%%%%%%%%%%%%%%%%%%%%%%%%%%%%%%%%%%%%%%
%End introduction
%%%%%%%%%%%%%%%%%%%%%%%%%%%%%%%%%%%%%%%%%%%%%%%%%%%%%%%%%%%%%%%%%%%%%%%%%%%%%

%%%%%%%%%%%%%%%%%%%%%%%%%%%%%%%%%%%%%%%%%%%%%%%%%%%%%%%%%%%%%%%%%%%%%%%%%%%%%
%The system
%%%%%%%%%%%%%%%%%%%%%%%%%%%%%%%%%%%%%%%%%%%%%%%%%%%%%%%%%%%%%%%%%%%%%%%%%%%%%

\begin{figure}
\begin{center}
\includegraphics[width=0.9\columnwidth]{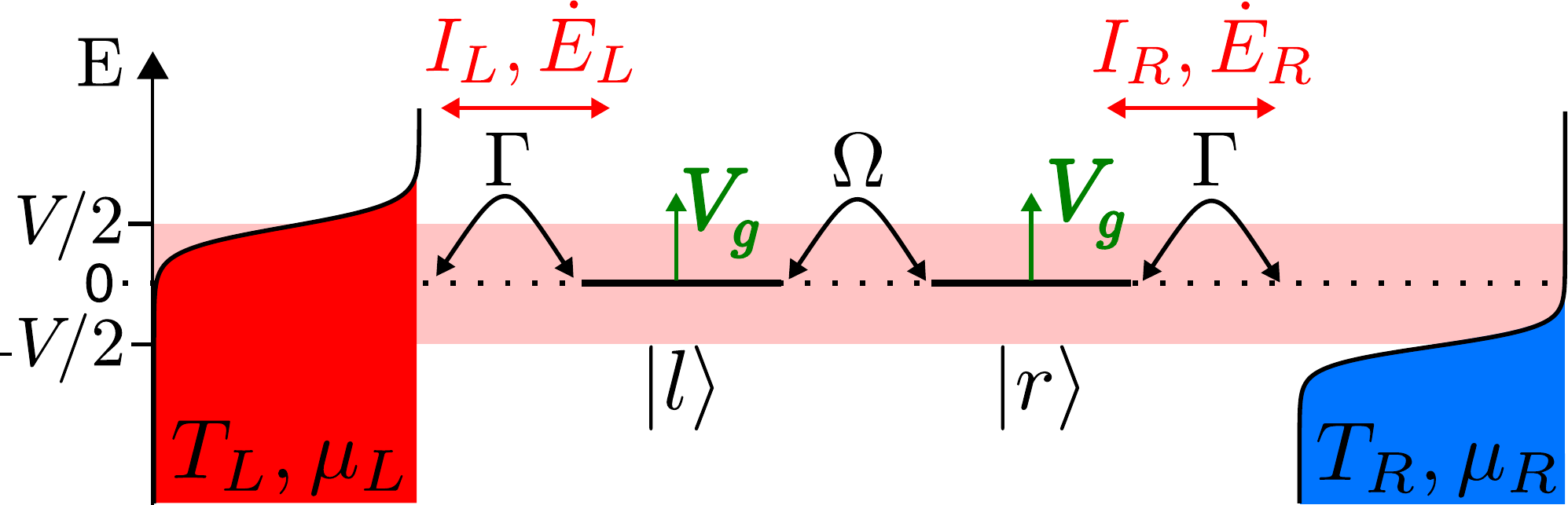}
\caption{(Color online) Spin-polarized double-dot structure. The energy of the dot states is shifted by a gate voltage $V_g$. Both dots are coupled to each other ($\Omega$) and {to} one lead each ($\Gamma$). The two leads are described as electron reservoirs at distinct temperatures $T_\ell$ and chemical potentials $\mu_\ell$, where {$\ell \in \{L, R\}$}. A difference in either of the two parameters between the two leads can result in the flow of particle {and} energy currents, $I_{\ell}$ and $\dot{E}_{\ell}$, respectively. An interaction energy $U$ can be {present} for the double-occupied state (not shown).}
\label{Figure_scenario}
\end{center}
\end{figure}

\begin{figure*}[t]
\begin{center}
\includegraphics[width=\textwidth]{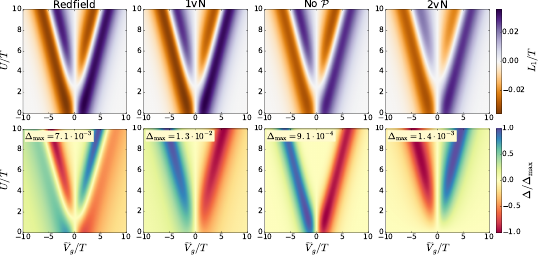}
\caption{(Color online) Off-diagonal Onsager coefficient $L_1$ (top row) and the deviation from Onsager's theorem, $\Delta=L_1' - L_1$ (bottom row), as functions of the charging energy $U$ and the gate voltage $\widetilde{V}_{g}=V_{g}+U/2$, calculated using the indicated approaches. The maximum deviation $\Delta_\textrm{max}$ uncovers significant violations of the Onsager relation \eqref{Onsager_finalExpression} for the first-order Redfield and 1vN approaches. Neglecting the {principal-value} integrals (``No $\mathcal{P}$'') or including second-order contributions ({``2vN''}) {significantly} reduces the violation. In all calculations we set $\Gamma=2\Omega=T/2$.}
\label{Figure_Onsager}
\end{center}
\end{figure*}

\section{Model}

We consider the example of a fully spin-polarized, serial double-dot structure coupled symmetrically to source ($L$) and drain ($R$) leads, as shown in \figureref{Figure_scenario}. The Hamiltonian is \cite{GurvitzPRB1996,GurvitzPRB1998,NovotnyEPL2002,LevyEPL2014} $H = H_\te{dot} + H_\te{leads} + H_\te{hyb}$, where
\begin{subequations}
\begin{align}
\label{Hamiltonian_dot}
H_\te{dot} &= V_g\,(\dd_{l}\dan_{l} + \dd_{r}\dan_{r})
  - \Omega\, (\dd_{l}\dan_{r} + \dd_{r}\dan_{l}) \nonumber \\
&\quad{}+ U\, \dd_{l}\dan_{l}\dd_{r}\dan_{r}, \\
\label{Hamiltonian_leads}H_\te{leads} &= \sum\limits_{\ell\f{k}}
  E_{\ell\f{k}}^{\phantom{\dag}}\, \cd_{\ell\f{k}} \can_{\ell\f{k}}, \\
\label{Hamiltonian_hyb} H_\te{hyb} &= \sum\limits_{\f{k}}
  t\, (\dd_{l}\can_{L\f{k}}+\dd_{r}\can_{R\f{k}}) + \te{{H.c.}}
\end{align}
\end{subequations}
Here, $\cd_{\ell\f{k}}$ creates an electron with quantum {numbers} $k$ in the lead {$\ell\in \{L,R\}$} and $\dd_i$ creates an electron in the dot $i\in\{l,r\}$. The coupling between the left dot ($l$) and the right dot ($r$) is given by the hybridization $\Omega$, while the level positions are controlled by the gate voltage $V_{g}$. Additionally, there is a charging energy $U$ when both {dots} are occupied. The energy dispersion in the leads is given by $E_{\ell\f{k}}$ and the electrons can tunnel between dots and leads with a tunneling amplitude $t$. The latter is {expressed in terms of} the tunneling rate $\Gamma=2\pi\sum_k t^2\,\delta(E-E_{\ell k})$, which is assumed to be independent of the energy $E$ (wide-band limit).  We also assume that the leads are thermalized according to a Fermi-Dirac occupation function $f_\ell(E)=[e^{(E-\mu_{\ell})/T_\ell}+1]^{-1}$ with different temperatures $T_{L/R}$ and chemical potentials $\mu_{L/R}$ for the two leads.

In the presence of a bias $V=\mu_{L}-\mu_{R}$ or a temperature difference between the leads, $\Delta{T}=T_{L}-T_{R}$, particle and energy currents flow. The currents are defined as
\begin{subequations}
\begin{align}
I_{\ell}(t) = -\tfrac{\partial}{\partial t}\avgs{N_{\ell}} &= -i\,\avgs{[H, N_{\ell}]}
\label{general_currentDef},
\\
\dot{E}_{\ell}(t) = -\tfrac{\partial}{\partial t}\avgs{H_{\ell}} &= -i\,\avgs{[ H, H_{\ell}]},
\label{general_heatcurrentDef}
\end{align}
\end{subequations}
where $N_{\ell}=\sum_{k}\cd_{\ell k}\can_{\ell k}$ and $H_{\ell}=\sum_{k}E_{\ell k}^{\phantom{\dag}}\,\cd_{\ell k}\can_{\ell k}$. Throughout this paper we consider the stationary state, where particle and energy conservation require $I_{L}=-I_{R}$ and $\dot{E}_{L}=-\dot{E}_{R}$, respectively. In practice, we use the currents emanating from the left lead, $I=I_{L}$ and $\dot{E}=\dot{E}_{L}$. In linear response to an applied bias $V$ and a temperature difference $\Delta{T}$, these currents can be expressed as
\begin{equation}
\begin{pmatrix}
I \\ \dot{E}
\end{pmatrix}
=
\begin{pmatrix}
L_{0} & L_{1} \\
L_{1}' & L_{2}
\end{pmatrix}
\begin{pmatrix}
V \\ \Delta{T}/{T}
\end{pmatrix}.
\label{OnsagerMatrix}
\end{equation}
To determine $L_{1}\approx TI/\Delta{T}$ and $L_{1}'\approx\dot{E}/V$ in linear response, we calculate the particle current
for $\mu_L=\mu_R = 0$ and $T_{L/R} = T \pm \Delta T / 2$ using $\Delta T=0.01\,T
$. The energy current is determined for $\mu_{L/R}=\pm V/2$ and $T_L=T_R$
using $V=0.01\,T$.
Onsager's theorem predicts that the off-diagonal coefficients $L_{1}=T\,(\pd I/\pd\Delta{T})_{\Delta{T}=0}$ and $L_{1}'=(\pd \dot{E}/\pd V)_{V=0}$
are equal, i.e., {that}
\begin{equation}\label{Onsager_finalExpression}
\Delta\equiv L_{1}'-L_{1} {=0} .
\end{equation}
Note that for our choices of $\mu_L$ and $\mu_R$, the energy and the heat current are identical within linear response. This can be seen by expanding the heat current in powers of $\Delta T$ and $\Delta \mu$, where the only contribution to $L_1'$ of order unity is due to the energy current (see Appendix A). %\cite{SupplementalMaterial}.

%%%%%%%%%%%%%%%%%%%%%%%%%%%%%%%%%%%%%%%%%%%%%%%%%%%%%%%%%%%%%%%%%%%%%%%%%%%%%
% End the system
%%%%%%%%%%%%%%%%%%%%%%%%%%%%%%%%%%%%%%%%%%%%%%%%%%%%%%%%%%%%%%%%%%%%%%%%%%%%%

\section{Numerical results}

%%%%%%%%%%%%%%%%%%%%%%%%%%%%%%%%%%%%%%%%%%%%%%%%%%%%%%%%%%%%%%%%%%%%%%%%%%%%%
% Approaches
%%%%%%%%%%%%%%%%%%%%%%%%%%%%%%%%%%%%%%%%%%%%%%%%%%%%%%%%%%%%%%%%%%%%%%%%%%%%%
In this work, we evaluate the coefficients $L_{1}'$ and $L_{1}$ using the \textit{Pauli}, \textit{Redfield}, \textit{first-order von Neumann} (1vN), and \textit{second-order von Neumann} (2vN) MEs. All the mentioned first-order approaches can be derived from the Wangsness-Bloch-Redfield (WBR) equation \cite{BlochPR1957,RedfieldIBM1957, RedfieldAdvMagnOptReson1965,WangsnessPR1953,HarbolaPRB2006}. Projecting this equation onto dot eigenstates gives equations for the elements of the reduced density matrix. The \emph{Redfield approach} uses the resulting equations for all diagonal elements (populations) and those off-diagonals (coherences) that link states with equal charge. All other off-diagonals {decay} rapidly due to superselection rules \cite{ZurekPRD1982}. One can derive a similar set of equations using a different Markov approximation, the so-called \emph{first-order von Neumann} (1vN) approach \cite{PedersenPRB2007}. In both approaches, we can obtain negative diagonal elements, a known problem of WBR-type equations \cite{BreuerBook2006}. Another common feature of both approaches is the appearance of principal-value integrals over the lead states, which are often neglected \cite{GardinerBook2000,HusseinPRB2014,GoldozianSciRep2016}. For our system, doing so yields the same steady-state currents in both Redfield and 1vN approaches, referred to as ``No $\mathcal{P}$.'' Neglecting the off-diagonal elements of the reduced density matrix reduces both approaches to the Pauli ME.
%%%%%%%%%%%%%%%%%%%%%%%%%%%%%%%%%%%%%%%

The 1vN equations can also be derived in a hierarchical approach where all processes corresponding to single-electron or single-hole excitations in the leads are taken into account. Similarly, the 2vN approach includes all processes of up to two electron and hole excitations. This extension captures level broadening as well as coherent effects and cotunneling \cite{PedersenPRB2005}. %Details on all approaches are given in the Supplemental Material \cite{SupplementalMaterial}.

%%%%%%%%%%%%%%%%%%%%%%%%%%%%%%%%%%%%%%%%%%%%%%%%%%%%%%%%%%%%%%%%%%%%%%%%%%%%%
% End Approaches
%%%%%%%%%%%%%%%%%%%%%%%%%%%%%%%%%%%%%%%%%%%%%%%%%%%%%%%%%%%%%%%%%%%%%%%%%%%%%

%%%%%%%%%%%%%%%%%%%%%%%%%%%%%%%%%%%%%%%%%%%%%%%%%%%%%%%%%%%%%%%%%%%%%%%%%%%%%
% General Results
%%%%%%%%%%%%%%%%%%%%%%%%%%%%%%%%%%%%%%%%%%%%%%%%%%%%%%%%%%%%%%%%%%%%%%%%%%%%%

\begin{figure}[t]
\begin{center}
\includegraphics[width=\columnwidth]{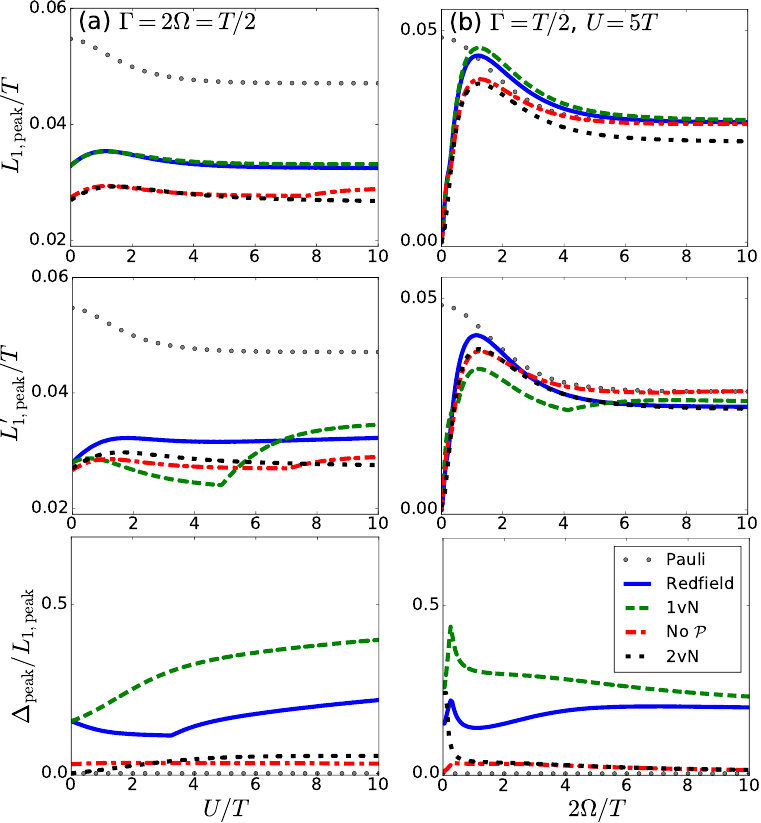}
\caption{(Color online) {Dependence} of the peak values
$L_{1,\peak}$, $L_{1,\peak}'$ of the Onsager coefficients and their difference $\Delta_{\peak}$ on $U$ (left column) and $\Omega$ (right column) upon varying the gate voltage $V_g$ for given parameters $T$, $\Omega$, $U$. For increasing $U$, the deviation $\Delta_{\peak}$ saturates in all approaches. Increasing $\Omega$ leads to saturated $\Delta_{\peak}$ for Redfield and 1vN approaches and decaying $\Delta_{\peak}$ for the 2vN and ``No $\mathcal{P}$'' approaches.
}
\label{curves_omega}
\end{center}
\end{figure}

Figure~\ref{Figure_Onsager} compares the results of the different approaches. The calculated values for $L_1$ agree fairly well for all approaches shown. The 2vN results are slightly more smeared-out due to the inclusion of level broadening, which is an effect of second order in $\Gamma$. The first-order Redfield and 1vN approaches exhibit significant violations of the Onsager relation \eqref{Onsager_finalExpression}, where the deviations $\Delta$ reach 22\% and 40\% of the maximum of $L_1$, respectively. Neglecting the principal-value integrals in either of the two approaches (No $\mathcal{P}$) reduces the violation to 3\%. The 2vN approach satisfies the Onsager relation for the non-interacting system $U=0$ but moderate violations, up to 5\%, arise for $U>0$. All results are antisymmetric with respect to the line $V_g=-U/2$ due to electron-hole symmetry. Results for the Pauli ME are not shown since Eq.\ \eqref{Onsager_finalExpression} is satisfied exactly for this approach. For a quantitative comparison of the different approaches, we plot the peak values $L_{1,\peak}=\textrm{max}_{V_{g}}\abs{L_{1}(V_{g})}$ and $\Delta_{\peak}=\textrm{max}_{V_{g}}\abs{\Delta(V_{g})}$
in \figurename~\ref{curves_omega}.

Column (a), which is based on the data displayed in \figurename~\ref{Figure_Onsager}, shows the dependence on $U$. In this figure, we also show the predictions of the Pauli ME, which satisfy Onsager's theorem but yield completely different values for the coefficients $L_1$ and $L_{1}'$ compared to the other approaches, as discussed further below.

We see that the Redfield and 1vN approaches give similar results for $L_1$ but very different behaviors of $L_1'$. As a result, the two approaches do not predict the same difference $\Delta_\te{peak}$. We also note that the predictions of these two approaches are very different from those of the other ones. For all methods, $\Delta_\te{peak}$ remains finite and saturates for large $U$.
Column (b) of \figurename~\ref{curves_omega} shows the dependence of $\Delta_{\peak}$ on the dot hybridization $\Omega$ for $U=5T$. Again, the Redfield and 1vN {approaches} essentially differ in $L_{1}'$ and show {a strong violation} of the Onsager relation for all values of $\Omega$. The coefficients $L_1$ and $L_{1}'$ drop to zero for vanishing $\Omega$, i.e., for decoupled dots, in all approaches except for the Pauli ME. This is a general failure of the Pauli ME if the level spacing $2\Omega$ becomes smaller than $\Gamma$, see also the discussion in Ref.~\onlinecite{GoldozianSciRep2016} and references therein. For large $\Omega$, the Pauli ME agrees with the first-order approaches only if the principal-value parts are neglected.
%%%%%%%%%%%%%%%%%%%%%%%%%%%%%%%%%%%%%%%%%%%%%%%%%%%%%%%%%%%%%%%%%%%%%%%%%%%%%
% End General Results
%%%%%%%%%%%%%%%%%%%%%%%%%%%%%%%%%%%%%%%%%%%%%%%%%%%%%%%%%%%%%%%%%%%%%%%%%%%%%

\section{Analytical results for $U=0$}

%%%%%%%%%%%%%%%%%%%%%%%%%%%%%%%%%%%%%%%%%%%%%%%%%%%%%%%%%%%%%%%%%%%%%%%%%%%%%
% U=0
%%%%%%%%%%%%%%%%%%%%%%%%%%%%%%%%%%%%%%%%%%%%%%%%%%%%%%%%%%%%%%%%%%%%%%%%%%%%%

In order to gain more insight, we consider the non-interacting case, $U=0$, where exact results for the currents can be obtained using the transmission formalism \cite{CaroliJPhysCSolidStatePhys1971,ButtikerPRB1985,ButtikerPRL1986,DattaBook1995}. A straightforward calculation
in Appendix D gives
%\cite{SupplementalMaterial} gives
%
\begin{subequations}\label{LB_currents}
\begin{align}\label{LB_current}
I_{\mathcal{T}} &= \frac{\Gamma}{4\pi}\Re\big[b\,(\Psi_{L-}-\Psi_{R-})
  - b^{*}\,(\Psi_{L+}-\Psi_{R+})\big], \\
\label{LB_heatcurrent}
\dot{E}_{\mathcal{T}} &= \frac{\Gamma}{4\pi}\Re\big[b\,(V_g-\Omega-i\Gamma/2)
  (\Psi_{L-}-\Psi_{R-}) \nonumber \\
&\quad {}-b^{*}\,(V_g+\Omega-i\Gamma/2)(\Psi_{L+}-\Psi_{R+})\big],
\end{align}
\end{subequations}
where $b=(i+\gamma)/(1+\gamma^2)$, $\gamma=\Gamma/(2\Omega)$, and
\begin{equation}\label{psigam}
\Psi_{\ell\pm}=\Psi\left( \frac{1}{2}
  + \frac{\mu_\ell - (V_g \pm \Omega-i\Gamma/2)}
    {{2 \pi i\,} T_\ell}\right),
\end{equation}
with the digamma function $\Psi(z)$. This {gives} the off-diagonal Onsager coefficients
\begin{align}
L_{1,\mathcal{T}} = L_{1,\mathcal{T}}'
&= \frac{\Gamma}{4\pi} \Re \bigg[
b\,\Psi_-' \frac{V_g - \Omega - i \Gamma / 2}{{2 \pi i\,} T} \nonumber \\
&\quad {}-b^*\,\Psi_+'  \frac{V_g + \Omega - i \Gamma / 2}{{2\pi i\,} T} \bigg] ,
\label{TF_conductance}
\end{align}
where $\Psi_\pm'$ is the derivative of the digamma function in Eq.~\eqref{psigam} at $T_L=T_R=T$ and $\mu_L=\mu_R=\mu$. Thus, the Onsager relation \eqref{Onsager_finalExpression} is satisfied exactly. The exact results for $U=0$ allow for a comparison to the analytical expressions obtained in the ME approaches (see Appendix E). %\cite{SupplementalMaterial}.

The 2vN approach gives the exact stationary current for non-interacting systems. This was shown analytically in Ref.~\onlinecite{KarlstromJPA2013} and explains the vanishing of $\Delta$ calculated within the 2vN approach for $U=0$, as seen in Fig.~\ref{Figure_Onsager}.

The Pauli ME gives the steady-state currents
\begin{subequations}\label{Pauli_currents}
\begin{align}
I_{\te{P}} &= \frac{\Gamma}{4}\, (g_{+}+ g_{-}), \label{Pauli_currentU0}\\
\dot{E}_{\te{P}} &=  \frac{\Gamma}{4}\, \big[(V_g+\Omega) g_{+} + (V_g-\Omega)g_{-}\big] \label{Pauli_heatcurrentU0},
\end{align}
\end{subequations}
where $g_{\pm}=f_{L}(V_{g}\pm\Omega) - f_{R}(V_{g}\pm\Omega)$. Clearly, the two currents satisfy Eq.~\eqref{Onsager_finalExpression}. However, for vanishing coupling between the dots, $\Omega\rightarrow0$, the currents stay finite. As noted, this is unphysical and thus contradicts the exact result, Eqs.~\eqref{LB_currents}, where $I, \dot{E}\rightarrow 0$ \cite{SchultzPRB2009}.

Both the Redfield and 1vN approaches give
\begin{subequations}\label{R_currents}
\begin{align}\label{R_current}
I_{\te{Red}} &= \frac{\Gamma}{4\pi} \Re\big[b\,(\psi_{L-}-\psi_{R-})
  - b^{*}\,(\psi_{L+}-\psi_{R+})\big], \\
\label{R_heatcurrent}
\dot{E}_{\te{Red}} &= \frac{\Gamma}{4\pi}\Re\big[b\,(V_g-\Omega-i\Gamma/2)
  (\psi_{L-}-\psi_{R-})  \nonumber \\
&\quad {}-b^{*}\,(V_g+\Omega-i\Gamma/2)(\psi_{L+}-\psi_{R+})\big],
\end{align}
\end{subequations}
where $\psi_{\ell\pm}$ is the expression $\Psi_{\ell\pm}$ from Eq.~\eqref{psigam} with $\Gamma=0$. This is the only difference to the exact transmission result \eqref{LB_currents}. This yields
\begin{equation}\label{deltared}
\Delta_{\te{Red}}=L_{1,\te{Red}}'-L_{1,\te{Red}}=
-\frac{\Gamma^2}{8\pi^2 T}\Re \left( b  \psi_{-}' - b^*  \psi_{+}'\right) .
\end{equation}
Here, $\psi_{\pm}' = \Psi'(1/2 - (V_g \pm \Omega)/2 \pi i\, T)$
is the derivative of the digamma function. The main finding is that the violation of the Onsager relation is proportional to $\Gamma^2$ and thus goes beyond the terms of first order in $\Gamma$ that are fully taken into account in the Redfield and 1vN approaches. Indeed, the terms $-i\Gamma/2$ in Eq.~(\ref{R_heatcurrent}) provide higher order terms in $\Gamma$, and the finite value of {$\Delta_{\te{Red}}$} can be traced back to precisely these
terms \footnote{Further contributions to higher orders in $\Gamma$ result from the factors $b$. However, the violation of Onsager's theorem does not occur if the terms $-i\Gamma/2$ are neglected, while the higher-order terms in $b$ do not affect this.}. This also suggests that the energy current is more problematic than the particle current in these approaches.

Finally, the ``No $\mathcal{P}$'' variant gives
\begin{subequations}\label{R_noP_currents}
\begin{align}
{I_{\te{No}\:\mathcal{P}}} &= \frac{\Gamma}{4(1+\gamma^2)}\, (g_{+} + g_{-}), \label{Redfield_currentU0}\\
{\dot{E}_{\te{No}\:\mathcal{P}}} &=\frac{\Gamma}{4(1+\gamma^2)}\,
  \big[(V_g+\widetilde{\Omega})\,g_{+}
  + (V_g-\widetilde{\Omega})\,g_{-}\big]
\label{Redfield_heatcurrentU0},
\end{align}
\end{subequations}
where $\widetilde{\Omega} = \Omega\,(1 + \gamma^2)$. In the limit $\Omega\gg\Gamma$, i.e., $\gamma\to 0$, this agrees with the result (\ref{Pauli_currents}) of the Pauli ME. In the opposite case $\Omega\ll\Gamma$, i.e., $\gamma\to \infty$, the currents drop as expected, curing the failure of the Pauli approach. It is worth noting that the energy carried by the two resonances at $V_g \pm \Omega$ is shifted to $V_g \pm \widetilde{\Omega}$. This shift results in a difference between $L_{1,\te{No}\:\mathcal{P}}'$ and $L_{1,\te{No}\:\mathcal{P}}$,
\begin{align}\label{deltanop}
{\Delta_{\te{No}\:\mathcal{P}}
= L_{1,\te{No}\:\mathcal{P}}'-L_{1,\te{No}\:\mathcal{P}}}
= -\frac{\Gamma^3}{16T\widetilde{\Omega}}\, (f_+' - f_-') ,
\end{align}
where $f'_{\pm} =f'(V_{g}\pm\Omega)$ is the derivative of the Fermi function for $\mu_\ell=0$ and $T_\ell=T$. This is of third order in $\Gamma$, an improvement in comparison to $\Delta_{\te{Red}}$.

%%%%%%%%%%%%%%%%%%%%%%%%%%%%%%%%%%%%%%%%%%%%%%%%%%%%%%%%%%%%%%%%%%%%%%%%%%%%%
% End U=0
%%%%%%%%%%%%%%%%%%%%%%%%%%%%%%%%%%%%%%%%%%%%%%%%%%%%%%%%%%%%%%%%%%%%%%%%%%%%%

%%%%%%%%%%%%%%%%%%%%%%%%%%%%%%%%%%%%%%%%%%%%%%%%%%%%%%%%%%%%%%%%
% Gamma scaling
%%%%%%%%%%%%%%%%%%%%%%%%%%%%%%%%%%%%%%%%%%%%%%%%%%%%%%%%%%%%%%%%

\begin{figure}[t]
\centering
\includegraphics[width=\columnwidth]{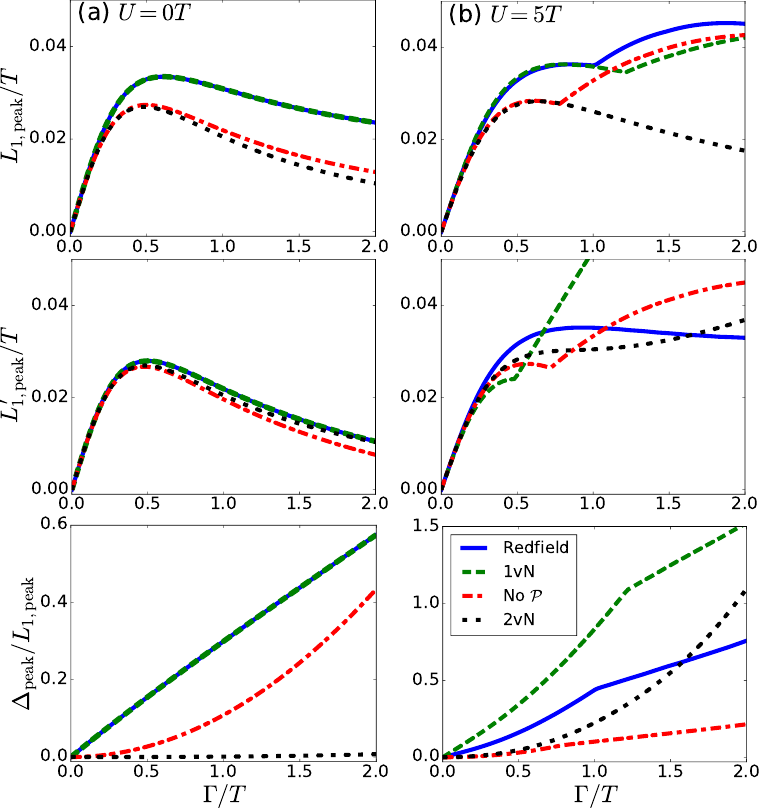}
\caption{(Color online) $\Gamma$ dependence of the peak values $L_{1,\peak}$, $L_{1,\peak}'$, and $\Delta_{\peak}$ for $U=0$ (left column) and $U=5T$ (right column) for $2\Omega=T/2$. The Pauli ME (not shown) gives $L_1=L_{1}'\propto \Gamma$ with $L_{1}$ and $L_{1}'$ having the same slope as in all other approaches at $\Gamma=0$ . For the Redfield and 1vN approaches the scaling of $\Delta_{\peak}/L_{1,\peak}$ is of order $\Gamma$ and for ``No $\mathcal{P}$'' and 2vN approaches the scaling is of order $\Gamma^2$. However, for $U=0$ Onsager's theorem is satisfied in the 2vN approach as the conductances agree with the exact transmission formalism, Eq.~\eqref{TF_conductance}.
   }\label{curves_gamma}
\end{figure}

\section{Scaling with $\Gamma$}

\begin{figure}[t]
\centering
\includegraphics[width=0.8\columnwidth]{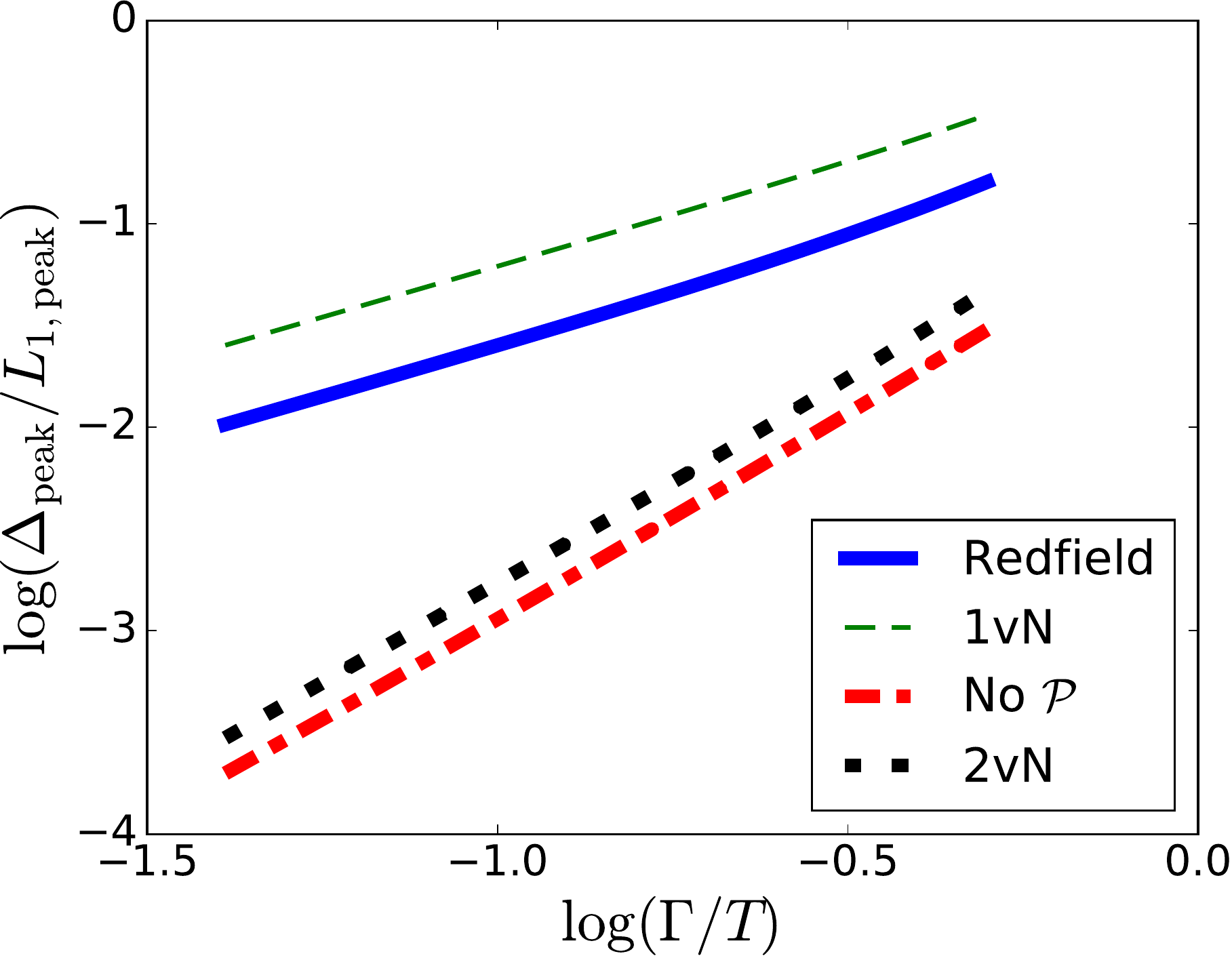}
\caption{(Color online) Dependence of $\Delta_{\peak}/L_{1,\peak}$ on $\Gamma$ shown in a {double logarithmic} plot. The figure corresponds to Fig.~4b zoomed into the region of small $\Gamma/T$. The values of parameters are $U=5T$ and $2\Omega=T/2$.}
\label{curves_gamma_sup}
\end{figure}

The analytical results for $U=0$ of the previous section show that deviations from the Onsager relation are of order $\Gamma^2$ for the Redfield and 1vN approaches and of order $\Gamma^3$ if the principal-value integrals are neglected. Numerical solutions for the various ME approaches, which exhibit the dependence on $\Gamma/T$, are shown in Fig.~\ref{curves_gamma}. For $U=0$, column (a), the numerical solutions of the Redfield, 1vN, and No $\mathcal{P}$ methods fully agree with the analytical results of Eqs.~\eqref{deltared} and \eqref{deltanop}. Redfield and 1vN provide identical results and {$\Delta_{\te{Red}}$} scales as $\Gamma^2$ for small $\Gamma$. The 2vN approach satisfies the Onsager relation, as the currents agree with the exact result, while $\Delta_{\te{No}\:\mathcal{P}}$ scales as $\Gamma^3$. For an interacting system with $U=5T$, column (b), the deviations $\Delta$ scale with small $\Gamma$ like for $U=0$. Here, the 2vN approach is no longer exact and violates the Onsager relation but the order of deviation $\Delta_{\te{2vN}}$ scales as $\Gamma^3$, i.e., these are terms of higher order than the second-order perturbation expansion in the coupling.

Finally, we zoom into the region of small $\Gamma/T$ {in} Fig.~\ref{curves_gamma} to demonstrate the scaling behavior of $\Delta_{\peak}/L_{1,\peak}$ more clearly. A double logarithmic plot of this region is presented in Fig.~\ref{curves_gamma_sup}. We see that for the Redfield and 1vN approaches {the scaling follows} $\Delta_{\peak}/L_{1,\peak}\sim\Gamma^1$ while for {the} ``No $\mathcal{P}$'' and 2vN approaches we observe $\Delta_{\peak}/L_{1,\peak}\sim\Gamma^2$. Since $L_{1,\peak}$ scales with $\Gamma$, the resulting dependence of the violation of Onsager's theorem, $\Delta_{\peak}$, is $\Gamma^2$ and $\Gamma^3$, respectively.

%%%%%%%%%%%%%%%%%%%%%%%%%%%%%%%%%%%%%%%%%%%%%%%%%%%%%%%%%%%%%%%%
% End Gamma scaling
%%%%%%%%%%%%%%%%%%%%%%%%%%%%%%%%%%%%%%%%%%%%%%%%%%%%%%%%%%%%%%%%

\section{Conclusions}

%%%%%%%%%%%%%%%%%%%%%%%%%%%%%%%%%%%%%%%%%%%%%%%%%%%%%%%%%%%%%%%%
%CONCLUSION
%%%%%%%%%%%%%%%%%%%%%%%%%%%%%%%%%%%%%%%%%%%%%%%%%%%%%%%%%%%%%%%%
In conclusion, we have shown that MEs that take into account coherences generically fail to satisfy Onsager's theorem. For small coupling to the leads, the deviations scale as a power of $\Gamma$ that is higher than the order of perturbation theory in the respective approach. In first-order approaches, such as Redfield {and} 1vN, the deviations scale as $\Gamma^2$. For thermoelectric systems, this restricts the applicability of popular first-order approaches to the weak-coupling limit $\Gamma\ll T$, even if the particle currents frequently exhibit good results up to moderate couplings. For our model, the violation of Onsager's theorem is pushed to a higher order in $\Gamma$ if the occurring principal-value integrals are neglected. It should be noted that they are required to catch essential physics like level energy renormalization in some systems \cite{KonigPRL2003,WunschPRB2005,PedersenPRB2007,SothmannPRB2010,MisiornyNatPhys2013}, meaning an ad-hoc neglect is not always justified. For the 2vN approach, the deviation is of order $\Gamma^3$ and provides an extended range of applicability.
Our results show that MEs formally contradict a well-established theory for systems out of equilibrium. However, the scaling behavior of the resulting deviations suggests that such approaches can still be confidently used to calculate transport for sufficiently weak coupling.

\begin{acknowledgments}
Financial support from the Swedish Research Council (VR), Grant No. 2012-4024, and NanoLund is gratefully acknowledged. C.\,T. acknowledges financial support by the Deutsche Forschungsgemeinschaft, through Research Unit FOR 1154 and Collaborative Research Center SFB 1143.
\end{acknowledgments}

\appendix

\section{\label{sec1:Redfield}Redfield and ``No $\mathcal{P}$'' approaches}

We obtain the Redfield approach {by} projecting the Wangsness-Bloch-Redfield (WBR)\cite{BlochPR1957,RedfieldIBM1957, RedfieldAdvMagnOptReson1965,WangsnessPR1953} equation
{for the reduced density operator $\rho(t)$ of the dot},
%\chred{[I DON'T KNOW WHAT} \verb+_\subdot+ \chred{IS GOOD FOR, SEEMS TO DO NOTHING]}
% that is true, it used to add a subscript "dot" but that was omitted at some point
\begin{align}\label{RedfieldFinal_Schr}
\pd_{t}\rho_\subdot(t) &= -i \bigl[H_\te{dot}, \rho_\subdot(t)\bigr]
 - \int\limits_{0}^{\infty} \dif{\tau}\, {\tr_E} \Bigl[ H_{\te{hyb}}, \nonumber \\
&\quad  \bigl[ e^{-iH_0\tau}H_{\te{hyb}}e^{iH_0 \tau},\rho_\subdot(t)\otimes
  {\rho^0_E} \bigr] \Bigr],
\end{align}
given here with all operators in the Schr\"{o}dinger picture,
onto the dot many-particle eigenstates $\ket{b}$, $\ket{b'}$.
{In terms of these states, the dot Hamiltonian reads as
$H_{\mathrm{dot}}=\sum_{b}E_b\, \ket{b}\bra{b}$.}
{Furthermore, $H_0 = H_\mathrm{dot} + H_\mathrm{leads}$, $\tr_E$ denotes the
trace over the degrees of freedom of the environment, i.e., the leads, and
$\rho^0_E$ is a density operator describing the leads in possibly separate thermal
equilibrium. The} hybridization Hamiltonian, expressed in the basis of {these}
states, becomes
\begin{align}
\label{hamT2}
H_{\mathrm{hyb}} &= \sum_{i,k\ell}\left(t_{i\ell}\,\dd_{i}\can_{\ell k}
  + {\mathrm{H.c.}} \right) \nonumber \\
&= \sum_{ab,k\ell}\left(T_{ba}^{\ell}\,\ket{b}\bra{a}\can_{\ell k}
  + {\mathrm{H.c.}} \right),
\end{align}
with {$T_{ba}^{\ell} = \sum_{i}t_{i\ell}\, \bra{b}\dd_{i}\ket{a}$ and
$T_{ab}^{\ell} = T^{\ell *}_{ba}$}.\cite{LetterConvention}
The reduced density matrix {has} to satisfy $\partial_t \rho_\subdot = 0$ for a steady-state solution. Explicitly, this gives the equations
\begin{align}\label{ss_RF}
%\frac{d}{dt} \rho_{bb'}
0 &= \rho_{bb'}\, (E_{b}-E_{b'}) \nonumber \\
&\quad {}+ \sum_{b''\ell}\rho_{bb''}\bigg[\sum_{a}\Gamma_{b''a,ab'}^{\ell}I_{b''a}^{\ell-}
  - \sum_{c}\Gamma_{b''c,cb'}^{\ell}I_{cb''}^{\ell+*} \bigg] \nonumber \\
&\quad {}+ \sum_{b''\ell}\rho_{b''b'}\bigg[\sum_{c}\Gamma_{bc,cb''}^{\ell}I_{cb''}^{\ell+}
  - \sum_{a}\Gamma_{ba,ab''}^{\ell}I_{b''a}^{\ell-*} \bigg] \nonumber \\
&\quad {}+ \sum_{aa'\ell}\rho_{aa'}\,\Gamma_{ba,a'b'}^{\ell}
  \big[I_{b'a'}^{\ell+*}-I_{ba}^{\ell+}\big] \nonumber \\
&\quad {}+ \sum_{cc'\ell}\rho_{cc'}\,\Gamma_{bc,c'b'}^{\ell}
  \big[I_{cb}^{\ell-*}-I_{c'b'}^{\ell-}\big],
\end{align}
which we combine with the normalization condition
\begin{equation}
\sum_{b}\rho_{bb}=1.
\label{rho_sideCondition}
\end{equation}
{In Eq.\ (\ref{ss_RF})}, the {tunneling-rate matrix} is defined as
\begin{equation}
\Gamma_{ba,a'b'}^{\ell}=2\pi\nu_{F}T_{ba}^{\ell}T_{a'b'}^{\ell},
\end{equation}
and we use
\begin{align}
2\pi I_{ba}^{\ell\pm} &= \mc{P}\!\int_{-D}^{D}\frac{\dif{E}\,f(\pm E)}{E-\xi_{ba}^{\ell}}
  -i\pi f(\pm \xi_{ba}^{\ell})\theta(D-\abs{\xi_{ba}^{\ell}}),
\label{ooxi} \\
\xi_{ba}^{\ell} &= E_{b}-E_{a}-\mu_{\ell},
\label{ooxi.2} \\
f(E)&=(\exp[E/T]+1)^{-1}.
\label{ooxi.3}
\end{align}
{We} have replaced $k$-sums using
{the approximation of}
a flat density of states, {i.e.}, $\sum_{k}\rightarrow \nu_{F}\int_{-D}^{D} \dif{E}$. Here, $\nu_{F}$ denotes the density of states at the Fermi {energy} and $2D$ is the symmetric bandwidth of the leads. We assume $D$ to be the largest energy scale in our system and take the limit $D\rightarrow\infty$. In this wide-band limit the results become {independent of bandwidth. Lastly, we have} assumed lead-electron dispersions of the form $E_{\ell k}=E_{k}+\mu_{\ell}$, with $E_{k}\in[-D,D]$, leading to Eq.~\eqref{ooxi}. %\cite{HaldanePRL1978}

The {particle and energy} currents, as defined in {Eqs.}\ \eqref{general_currentDef} and \eqref{general_heatcurrentDef}
{in the main text}, have the explicit {forms}
\begin{equation}
I_\ell = 2\Imag \bigg[ \sum\limits_{cbb'} \rho_{b'b}  \Gamma_{bc, cb'}^{\ell}
  I^{{\ell}+}_{cb'}
  - \sum\limits_{bcc'} \rho_{cc'} \Gamma^{\ell}_{c'b, bc} I^{\ell-}_{c'b} \bigg]
\label{RF_current}
\end{equation}
and
\begin{align}
\dot{E}_\ell
&= 2\Imag\bigg[ \sum\limits_{cbb'} \rho_{b'b}  \Gamma_{bc, cb'}^{\ell}
  (D + E_{cb'} I^{{\ell}+}_{cb'}) \nonumber\\
&\quad {}-\sum\limits_{bcc'} \rho_{cc'} \Gamma^{\ell}_{c'b, bc}
  (D + E_{c'b} I^{\ell-}_{c'b} ) \bigg] \nonumber \\
%%%
&= 2\Imag\bigg[\sum\limits_{cbb'} \rho_{b'b}  \Gamma_{bc, cb'}^{\ell}
  E_{cb'} I^{{\ell}+}_{cb'} \nonumber \\
&\quad {}- \sum\limits_{bcc'} \rho_{cc'} \Gamma^{\ell}_{c'b, bc}
  E_{c'b} I^{\ell-}_{c'b} \bigg],
\label{RF_energy}
\end{align}
where $E_{cb} = E_c - E_b$. The terms proportional to the bandwidth $D$ cancel
{since for every term $\rho_{bb'}\Gamma_{bc,cb'}$, the sum also contains its
complex conjugate,} and the sum of the two vanishes when taking the imaginary part.

{In principle,} we have to compare the charge current to the {\emph{heat} current instead of the energy current $\dot{E}_\ell$ in order to} check the validity of Onsager's theorem. {The heat current is
\begin{align}
\dot{Q}_{\ell} &= -\partial_t \avgs{H_{\ell}-\mu N_\ell} = -i\avgs{[ H, H_{\ell} - \mu_\ell N_\ell]} \nonumber \\
&= 2\Imag\bigg[ \sum\limits_{cbb'} \rho_{b'b}  \Gamma_{bc, cb'}^{\ell}
  \xi_{cb'}^\ell I^{{\ell}+}_{cb'} \nonumber \\
& \quad {}- \sum\limits_{bcc'} \rho_{cc'} \Gamma^{\ell}_{c'b, bc}
  \xi_{c'b}^\ell I^{\ell-}_{c'b}\bigg],
\label{RF_heat}
\end{align}
%\chred{[``$p$'' IS AN ODD SYMBOL FOR AN ENERGY, CALL THIS $q$ OR, BY STANDARD DEFINITION OF ENERGY RELATIVE TO CHEMICAL POTENTIAL, $\xi$?]}
}and has the energies $E_{cb}$ replaced by $\xi_{cb}^\ell = E_{cb} - \mu_\ell$ in comparison to Eq.~\eqref{RF_energy}. However, within linear response around $\mu=0$, the term proportional to $\mu_{L/R}=\pm V/2$ is of second order in the applied bias $V$ and thus irrelevant. Thus it is sufficient to calculate the energy current and compare it to the charge current.

{Finally, the} ``No $\mathcal{P}$'' approach corresponds to neglecting the principal-value {integral in} Eq.~\eqref{ooxi}.

%\vrule

{
\section{\label{sec:1vN}First-order and second-order von Neumann approaches}}

The usual derivation of \eqref{RedfieldFinal_Schr}, see, e.g., {Refs.\ \onlinecite{BreuerBook2006,TimmPRB2008}}, uses a Markov approximation to obtain a time-local equation. More precisely, this approximation is done for the reduced density matrix expressed in the interaction picture, {i.e.}, $\rho_{\subdotc \mathrm{I}}(t') \approx \rho_{\subdotc \mathrm{I}}(t)$ with the local time $t$. A different possible choice, $\rho_\subdot(t') \approx \rho_\subdot(t)$, leads to the \textit{first-order von Neumann} (1vN) approach.
In operator form, this Schr\"{o}dinger-picture Markov approximation leads to the equation
\begin{align}\label{1vN_IntAfterMarkov}
\pd_{t}\rho_\subdot(t) &= -i \bigl[ H_{\rm{dot}}, \rho_\subdot(t) \bigr]
  - \int\limits_{0}^\infty \dif{\tau}\, {\mathrm{tr}}_{\rm{E}} \Bigl[ H_{\mathrm{hyb}},
  \nonumber \\
&\quad e^{-iH_{0}\tau}\bigl[ H_{\rm{hyb}},  \rho_\subdot(t)\otimes \rho^{0}_{\mathrm{E}}   \bigr] e^{iH_{0}\tau} \Bigr].
\end{align}
{By} projecting Eq.~\eqref{1vN_IntAfterMarkov} onto dot eigenstates we obtain
\begin{align}\label{ss_1vN}
0 &= \rho_{bb'}\,(E_{b}-E_{b'}) \nonumber \\
&\quad {}+ \sum_{b''\ell}\rho_{bb''}\bigg[\sum_{a}\Gamma_{b''a,ab'}^{\ell}I_{ba}^{\ell-}
  -\sum_{c}\Gamma_{b''c,cb'}^{\ell}I_{cb}^{\ell+*} \bigg] \nonumber \\
&\quad {}+ \sum_{b''\ell}\rho_{b''b'}\bigg[\sum_{c}\Gamma_{bc,cb''}^{\ell}I_{cb'}^{\ell+}
  -\sum_{a}\Gamma_{ba,ab''}^{\ell}I_{b'a}^{\ell-*} \bigg] \nonumber \\
&\quad {}+ \sum_{aa'\ell}\rho_{aa'}\, \Gamma_{ba,a'b'}^{\ell}
  \big[I_{b'a}^{\ell+*}-I_{ba'}^{\ell+}\big] \nonumber \\
&\quad {}+ \sum_{cc'\ell}\rho_{cc'}\, \Gamma_{bc,c'b'}^{\ell}
  \big[I_{c'b}^{\ell-*}-I_{cb'}^{\ell-}\big].
\end{align}
%%%%%%%%%%%%%%%GEDIMINAS%%%%%%%%%%%
%&\begin{aligned}
%\frac{d}{dt} \rho_{bb'}=&\rho_{bb'}(E_{b}-E_{b'})\\
%+&\sum_{b''\ell}\rho_{bb''}\Big[\sum_{a}\Gamma_{b''a,ab'}^{\ell}I_{b''a}^{\ell-}-\sum_{c}\Gamma_{b''c,cb'}^{\ell}I_{cb''}^{\ell+*}\Big]\\
%+&\sum_{b''\ell}\rho_{b''b'}\Big[\sum_{c}\Gamma_{bc,cb''}^{\ell}I_{cb''}^{\ell+}-\sum_{a}
%\Gamma_{ba,ab''}^{\ell}I_{b''a}^{\ell-*}\Big]\\
%+&\sum_{aa'\ell}\rho_{aa'}\Gamma_{ba,a'b'}^{\ell}[I_{b'a'}^{\ell+*}-I_{ba}^{\ell+}]\\
%+&\sum_{cc'\ell}\rho_{cc'}\Gamma_{bc,c'b'}^{\ell}[I_{cb}^{\ell-*}-I_{c'b'}^{\ell-}].
%\end{aligned}
%%%%%%%%%%%%%%%%%%%%%%%%%%%%%%%%%%%%
Just as in the Redfield case, we solve Eq.~\eqref{1vN_IntAfterMarkov} under the constraint Eq.~\eqref{rho_sideCondition}. The definitions of $\Gamma$, $I$, and $f$ are the same as in {Eqs.~\eqref{ooxi}--\eqref{ooxi.3}}. As a result of the different Markov approximation, both the equation of motion {and} the currents are different. The explicit {expressions} for the particle and energy currents read {as}
\begin{align}
I_{\ell}
&= 2\Imag\bigg[\sum\limits_{cbb'} \rho_{b'b}  \Gamma_{bc, cb'}^{\ell} I^{{\ell}+}_{cb}
  - \sum\limits_{bcc'} \rho_{cc'} \Gamma^{\ell}_{c'b, bc} I^{\ell-}_{cb} \bigg],
\label{1vN_current} \\
\dot{E}_{\ell}
&= 2\Imag\bigg[\sum\limits_{cbb'} \rho_{b'b}  \Gamma_{bc, cb'}^{\ell} E_{cb} I^{{\ell}+}_{cb}
  \!-\! \sum\limits_{bcc'} \rho_{cc'} \Gamma^{\ell}_{c'b, bc} E_{cb} I^{\ell-}_{cb}\bigg],
% \nonumber \\
%&\quad {}- \sum\limits_{bcc'} \rho_{cc'} \Gamma^{\ell}_{c'b, bc} E_{cb} I^{\ell-}_{cb}\bigg],
\label{1vN_heat}
\end{align}
where {bandwidth-dependent} terms cancel by the same symmetry as in {Eq.~\eqref{RF_energy}}.
{The same remark about the heat and energy currents as in the Redfield approach applies to the 1vN approach.}

Comparison to the {Redfield equation} shows that the main difference between the two approaches is the energy assigned to processes involving coherences. As an example, consider the contribution of a single non-diagonal element $\rho_{b'b}~,b' \neq b$ to the energy current. In the Redfield approach {Eq.\ \eqref{RF_energy} contains
\begin{align}
\rho_{b'b}  \Gamma_{bc, cb'}^{\ell} E_{cb'} I^{{\ell}+}_{cb'}
&= \rho_{b'b}  \Gamma_{bc, cb'}^{\ell} E_{cb'}\bigg[
  \mc{P}\!\int_{-D}^{D}\frac{\dif{E}\,f(\pm E)}{E-\xi_{cb'}^{\ell}}
\nonumber \\
&\quad {}- i\pi f(\pm \xi_{cb'}^{\ell})\theta(D-\abs{\xi_{cb'}^{\ell}})
  \bigg].
\label{Redfield_Energy_contribution}
\end{align}
}In the 1vN approach, {we instead obtain
\begin{align}
\rho_{b'b}  \Gamma_{bc, cb'}^{\ell} E_{cb} I^{{\ell}+}_{cb}
&= \rho_{b'b}  \Gamma_{bc, cb'}^{\ell} E_{cb}\bigg[
  \mc{P}\!\int_{-D}^{D}\frac{\dif{E}\,f(\pm E)}{E-\xi_{cb}^{\ell}}
\nonumber \\
&\quad {}-i\pi f(\pm \xi_{cb}^{\ell})\theta(D-\abs{\xi_{cb}^{\ell}}) \bigg],
\label{1vN_Energy_contribution}
\end{align}
}which has all occurrences of the energy $E_{cb'}$ in Eq.~\eqref{Redfield_Energy_contribution} replaced by $E_{cb}$. A similar difference is found in the contributions of coherences to the equations of motion, Eqs.~\eqref{ss_RF} and \eqref{ss_1vN}.
{Note} that Eqs.~\eqref{ss_1vN}, \eqref{1vN_current}, and \eqref{1vN_heat} were originally derived in a framework that focuses on the number of excitations involved, in contrast to the {perturbative} derivation of the WBR equation. For a concise derivation, see the {supplementary information} of Ref.~\onlinecite{GoldozianSciRep2016}.

{The 1vN approach can be extended to the 2vN approach.}
%which is discussed in the following.
%
%\subsection{\label{sec:2vN}Second-order von Neumann approach}
%
A detailed derivation of {this} approach can be found in Ref.~\onlinecite{PedersenPRB2005} and a description of the numerical procedure used to solve the {resulting} equations is given in Appendix~A of Ref.~\onlinecite{KirsanskasPRB2016}. The source and drain leads are assumed to have a bandwidth of $2D$ with $E_{\ell k}=E_{k}\in[-D,D]$ and a constant density of states $\nu_{F}$ within this energy range. The chemical potentials are set to $\mu_{L/R}=\pm V/2$ for the energy current, and $\mu_\ell =0$ for the charge current. In the {present} work, we set the bandwidth to $2D=80T$ and {use} a lead-electron energy grid of $N=2^{13}$ points. %\chred{[REALLY? WHY SO VERY MANY?]}

%Additionally, we impose the normalisation condition for diagonal density matrix elements:
%
%
%
%We note that both the Redfield and 1vN approach neglect the broadening of the quantum-dot levels due to the dot-lead coupling. These effects could be included by a different treatment, see, e.g., Refs. [\onlinecite{PedersenPRB2005a,JinJCP2008,DordaPRB2014,ChenJPhysChemC2014}].

\vrule

\section{\label{sec:Pauli}Pauli master equation}

The Pauli master equation can be obtained from both the Redfield and 1vN approaches by neglecting all coherences $\rho_{bb'}$, $b\neq b'$. {By doing} so, we obtain
\begin{align}
\pd_{t}P_b &= \sum_{a\ell}\left[P_{a}\Gamma_{a\rightarrow b}^{\ell}f(+p_{ba}^{\ell})
  -P_{b}\Gamma_{b\rightarrow a}^{\ell}f(-p_{ba}^{\ell})\right] \nonumber \\
&\quad {}+ \sum_{c\ell}\left[P_{c}\Gamma_{c\rightarrow b}^{\ell}f(-p_{cb}^{\ell})
  -P_{b}\Gamma_{b\rightarrow c}^{\ell}f(+p_{cb}^{\ell})\right]
\end{align}
for the {probabilities} $P_{b}=\rho_{bb}$ {of dot many-particle states}. In this case, $\Gamma_{a\rightarrow b}^{\ell}=\Gamma_{ab,ba}^{\ell}=\Gamma_{b\rightarrow a}^{\ell}=\Gamma_{ba,ab}^{\ell}$. In the steady state, we again set $\partial_t P_b=0$ and solve the resulting equations under the constraint Eq.~\eqref{rho_sideCondition}. The steady-state currents then read {as}
\begin{align}\label{current}
I_{\ell} &= \sum_{bc} \big[P_{b}\Gamma_{b\rightarrow c}^{\ell}f(+p_{cb}^{\ell})
-P_{c}\Gamma_{c\rightarrow b}^{\ell}f(-p_{cb}^{\ell})\big] , \\
\label{heatcurrent12}
\dot{E}_{\ell} &= \sum_{bc}\big[P_{b}\Gamma_{b\rightarrow c}^{\ell} E_{cb}f(+p_{cb}^{\ell})
-P_{c}\Gamma_{c\rightarrow b}^{\ell} E_{cb} f(-p_{cb}^{\ell})\big].
\end{align}
%%%%%%%%%%%%%%%%%%%%%%%%%%%%%%%%%%%%%%%%%%%%%%%%%%%%%%%%%%%%%%%%%%%%%%%%%%%%%%%%%%%%%%%%

{\section{Transmission-function formalism}}

Here we present the main steps that lead to Eqs.~\eqref{LB_current} and \eqref{LB_heatcurrent} {in the main text}. The transmission-function formalism {\onlinecite{LandauerIBM1957, LandauerPhilosMag1970, ButtikerPRL1986} gives}
\begin{align}
I &= \frac{1}{2 \pi} \int\limits_{-D}^{D} \dif{E}\; \mathcal{T}(E)\, \bigl[ f_L(E) - f_R(E)\bigr],
\label{trans_current}
\\
\dot{E} &= \frac{1}{2 \pi} \int\limits_{-D}^{D} \dif{E}\; \mathcal{T}(E)\, E\, \bigl[ f_L(E) - f_R(E) \bigr]
\label{trans_heatcurrent}
\end{align}
for the currents. We evaluate these integrals in the {wide-band} limit $D\rightarrow \infty$. For the double-dot structure described in the main text, the transmission function reads {as}
\begin{align}\label{ddtrans}
\mathcal{T}(E)
%= \Gamma^2 \Omega^2 \biggl| \frac{1}{ \bigl( E-E_1-\Sigma_{11} \bigr) \bigl( E-E_2-\Sigma_{22} \bigr) }\biggr|^2
= \biggl| \frac{\Gamma/2}{E-E_{-} + i \Gamma /2 }- \frac{\Gamma/2}{E-E_{+}+i \Gamma/2}\biggr|^2,
\end{align}
where $E_{\pm}=V_{g}\pm\Omega$. This transmission function can be rewritten as
\begin{align}
\mathcal{T}(E) &= \frac{b}{E-E_{-} + i \Gamma /2} +\frac{b^*}{E-E_{-} - i \Gamma/2}
  \nonumber \\
&\quad {}- \frac{b^*}{E-E_{+}+i \Gamma/2} - \frac{b}{E-E_{+} - i \Gamma / 2},
\label{T_rewritten}
\end{align}
where
\begin{equation}
b = \frac{i+ \gamma}{1+\gamma^2},\quad {\gamma = \frac{\Gamma}{2\Omega}}.
\end{equation}
%\fi
For $D\gg \Gamma$ {and $D\gg E_{\pm}-\mu_{\ell}$}, the integrals in Eq.~\eqref{trans_current} can be expressed in terms of {the} digamma function $\Psi$~\cite{AbramowitzBook1972} using
\begin{equation}
\int\limits_{-D}^{D} \frac{\dif{E}\,f(E/T_\ell)}{E-(E_{\pm}-\mu_\ell)+i\Gamma/2}
%= \int\limits_{-R_\alpha}^{R^\alpha} dE\: \frac{f(x_\alpha)}{x_\ell+\varepsilon^\alpha_{\pm} + i \Gamma_\alpha/2}
\approx \Psi_{\ell\pm} -\ln \frac{D}{2\pi T_\ell}-i \frac{\pi}{2} ,
\label{integrationformula_text_LB}
\end{equation}
{where}
\begin{equation}
\Psi_{\ell\pm} \equiv \Psi\biggl( \frac{1}{2} + \frac{\mu_\ell - E_\pm + i \Gamma/2}{i 2 \pi T_\ell } \biggr).
\end{equation}
For the energy-current integrals in Eq.~\eqref{trans_heatcurrent}, we use
\begin{align}
\int\limits_{-D}^{D} &\dif{E} \,
  \frac{(E+\mu_\ell) f(E/T_\ell)}{E-(E_{\pm}-\mu_\ell)+i\Gamma/2} \nonumber \\
%&= D + (E_{\pm} -i\Gamma/2) \int\limits_{-D}^{D}\frac{dE\,f(E/T_\ell) }{E-(E_{\pm}-\mu_\ell)+i\Gamma/2}
%\nonumber
%\\
&\approx D + (E_\pm - i \Gamma /2 )\Bigl( \Psi_{\ell\pm} - \ln \frac{D}{2\pi T_\ell} - i \frac{\pi}{2}\Bigr).
\label{energycurrent_integral}
\end{align}
{From} the transmission function, Eq.~\eqref{T_rewritten}, and the identities {Eqs.~\eqref{integrationformula_text_LB}--\eqref{energycurrent_integral}} we get the particle and energy currents given by Eqs.~\eqref{LB_currents} in the main text.

\section{Detailed calculation for the Redfield and 1\lowercase{v}N approaches}

In this section we present more detailed calculations for the {double-dot} structure using the Redfield and 1vN approaches. {The dot Hamiltonian $H_\text{dot}$, Eq.\ \eqref{Hamiltonian_dot} in the main text, has four many-particle eigenstates,}
\begin{align}%\label{eigst}
&\ket{0}, &&E_{0}=0,\\
&\ket{1}=\dd_{1}\ket{0}, &&E_{1}=V_{g}-\Omega,\\
&\ket{1'}=\dd_{1'}\ket{0}, &&E_{1'}=V_{g}+\Omega,\\
&\ket{2}=\dd_{1'}\dd_{1}\ket{0}, &&E_{2}=2V_{g}+U,
\end{align}
where
\begin{equation}
\begin{pmatrix}\dan_{1} \\ \dan_{1'}\end{pmatrix}
=\frac{1}{\sqrt{2}}
\begin{pmatrix}\dan_{l}+\dan_{r} \\ \dan_{l}-\dan_{r}\end{pmatrix}.
\end{equation}
The {matrices of} the many-particle tunneling amplitudes {are}, in this
basis,
\begin{align}
{T}^{L}&=\frac{t}{\sqrt{2}}\begin{pmatrix}
0 & +1 &  +1 & 0 \\
+1 & 0 & 0 & +1 \\
+1 & 0 & 0 & -1 \\
0 & +1 & -1 & 0
\end{pmatrix},\\
{T}^{R}&=\frac{t}{\sqrt{2}}\begin{pmatrix}
0 & +1 &  -1 & 0 \\
+1 & 0 & 0 & -1 \\
-1 & 0 & 0 & -1 \\
0 & -1 & -1 & 0
\end{pmatrix}.
\end{align}
There are six non-vanishing elements of the {reduced density} matrix $\rho$, which we collect into the column vector {$\bm{\rho}={\big( \rho_{00}, \rho_{11}, \rho_{1'1'}, \rho_{22}, \rho_{11'}, \rho_{1'1} \big)^T}$}.
%\begin{pmatrix}
%\rho_{00} & \rho_{11} & \rho_{1'1'} & \rho_{22} & \rho_{11'} & \rho_{1'1}
%\end{pmatrix}^{T}.
Then the master equation {takes the form}
\begin{equation}\label{rholiouv}
\pd_{t}\bm{\rho}=\mathcal{L}\bm{\rho},
\end{equation}
with the Liouvillian $\mathcal{L}$.
\vfill

\begin{widetext}
 In the non-interacting case, $U=0$, after using Eq.~\eqref{ss_RF} for the Redfield approach and Eq.~\eqref{ss_1vN} for the 1vN approach{, respectively,} we obtain the Liouvillians
\renewcommand{\arraystretch}{1.25}
\begin{equation}
\mathcal{L}_\te{Red}= \frac{\Gamma}{2}
\begin{pmatrix}
-f_+ - f_-&\bar{f}_-& \bar{f}_+ & 0 & -C^{*} & -C
\\
f_- & -\bar{f}_- - f_+ & 0 & \bar{f}_+ & C^{*} &   C &
\\
f_+ & 0 & -\bar{f}_+ - f_-  & \bar{f}_- & C^{*} & C
\\
0& f_+ & f_- & -\bar{f}_+ - \bar{f}_- & -C^{*} & -C
\\
C & C & C & C &2i/\gamma-2&0
\\
C^{*} & C^{*} & C^{*} & C^{*} &0&-2i/\gamma-2
\end{pmatrix},% \\
\end{equation}
\begin{equation}
\mathcal{L}_\te{1vN}= \frac{\Gamma}{2}
\begin{pmatrix}
-f_+ - f_-&\bar{f}_-& \bar{f}_+ & 0 & -C & -C^{*}
\\
f_- & -\bar{f}_- - f_+ & 0 & \bar{f}_+ & C & C^{*}
\\
f_+ & 0 & -\bar{f}_+ - f_-  & \bar{f}_- & C & C^{*}
\\
0& f_+ & f_- & -\bar{f}_+ - \bar{f}_- & -C & -C^{*}
\\
C & C & C & C &2i/\gamma-2&0
\\
C^{*} & C^{*} & C^{*} & C^{*} &0&-2i/\gamma-2
\end{pmatrix}.
\end{equation}
\end{widetext}
Here, we {have} used the limit $D\rightarrow+\infty$ and the definitions
\begin{align}
\gamma &= \Gamma/2\Omega, \\
f_{\pm} &= 1+ \frac{1}{2\pi i}\big[(\psi_{L\pm}^{*}+\psi_{R\pm}^{*})
  -(\psi_{L\pm}+\psi_{R\pm})\big] \nonumber \\
&= f_{L}(V_{g}\pm\Omega)+f_{R}(V_{g}\pm\Omega), \\
\bar{f}_{\pm} &= 2-f_{\pm}, \\
C &= \frac{1}{2\pi i}[(\psi_{L+}^{*}-\psi_{R+}^{*})-(\psi_{L-}-\psi_{R-})], \\
\psi_{\ell \pm} &= \Psi\biggl( \frac{1}{2}
  + \frac{\mu_\ell - E_\pm}{i 2 \pi T_\ell } \biggr).
\end{align}
For the Redfield approach, the steady-state solution of Eq.~\eqref{rholiouv}, i.e., {the solution of} $\pd_{t}\bm{\rho}=0$ {satisfying} the normalization condition \eqref{rho_sideCondition}, reads {as}
%\begin{subequations}\label{sols}
\begin{align}
\label{sol0}\rho_{00}&=\frac{1}{4} \bar{f}_+ \bar{f}_- - \frac{1}{2}\Real{(C\rho_{1'1})}, \\
\label{sol1}\rho_{11}&=\frac{1}{4} f_- \bar{f}_+ + \frac{1}{2}\Real{(C\rho_{1'1})},\\
\label{sol1p}\rho_{1'1'}&=\frac{1}{4}f_+ \bar{f}_- + \frac{1}{2}\Real{(C\rho_{1'1})},
\end{align}\begin{align}
\label{sol2}\rho_{22}&=\frac{1}{4}f_+f_- - \frac{1}{2}\Real{(C\rho_{1'1})},\\
\label{sol1p1}\rho_{11'}&=\rho_{1'1}^{*}=\frac{i\gamma C}{2(1+i\gamma)}.
\end{align}
%\end{subequations}
%
The steady-state solution of the 1vN approach is obtained by replacing all terms $\Real{(C\rho_{1'1})}$ in Eqs.~{\eqref{sol0}--\eqref{sol2} by $\Real(C^{*}\rho_{1'1})$}. Since $\rho_{1'1}$ is the same in the two approaches, this change $C \rightarrow C^*$ corresponds to {changing} the energy assigned to $\rho_{1'1}$ from {$V_g + \Omega$} to {$V_g- \Omega$}. This is exactly what has been noted using Eqs.\ \eqref{Redfield_Energy_contribution}--\eqref{1vN_Energy_contribution}, i.e., the two approaches assigning different energies to processes involving coherences. In this case, the symmetric splitting of $2 \Omega$ around $V_g$ results in the symmetry between the solutions.
%In this case, the symmetry between $V_g \pm \Omega$ results in the symmetry of the solutions.

Inserting the solution \eqref{sol0}--\eqref{sol2} into Eqs.~\eqref{RF_current} and \eqref{RF_energy} gives the particle current, Eq.~\eqref{R_current}, and {the} energy current, Eq.~\eqref{R_heatcurrent}, of the main text. We note that in the Redfield approach
\begin{equation}
\Real{(C\rho_{1'1})}
  = \frac{1}{2}\frac{ \gamma^2 \abs{C}^2}{1+\gamma^2}\equiv 2\abs{\rho_{1'1}}^2,
\end{equation}
and hence {we} see that the inclusion of coherences corrects the diagonal elements by {terms proportional to} $\pm |\rho_{1'1}|^2$.
Lastly, the ``No $\mathcal{P}$'' result {of} a calculation in which principal-value integrals are neglected, is obtained by making the replacement $\psi_{\ell\pm}\rightarrow-i\pi f_{\ell}(V_{g}\pm\Omega)$, which yields % as can be seen from
\begin{align}
2C_{\mathrm{No} \ \!\mathcal{P}} &= f_{L}(V_{g}+\Omega)-f_{R}(V_{g}+\Omega)
  \nonumber \\
&\quad {}+f_{L}(V_{g}-\Omega)-f_{R}(V_{g}-\Omega) \nonumber \\
%\frac{1}{2\pi i}[(\psi_{L+}^{*}-\psi_{R+}^{*})-(\psi_{L-}-\psi_{R-})]
&\equiv g_{+}+g_{-}.
\end{align}
%
\begin{comment}
The currents then read
\begin{align}
I
&= - \frac{\Gamma}{4} \frac{1}{1+\gamma^2} \Bigl(g_+ + g_- - \gamma A \Bigr),
\\
\dot{E}
&=- \frac{\Gamma}{4} \frac{1}{1+\gamma^2} \Bigl( g_+(V_g+ \tilde{\Omega}) + g_- (V_g - \tilde{\Omega}) -  V_g \gamma A \Bigr).
\end{align}
which in Eq. \eqref{R_current} and Eq. \eqref{R_heatcurrent} were expressed using digamma functions.[QUOTE HERE] If we set $A=0$, we restore the results of the $``\te{no}~\mathcal{P}\te{``}$ calculation.
\end{comment}
%

%\bibliographystyle{apsrev4-1}
%\bibliography{refs_QuantTrans,refs_kevin,refs_gediminas}
%\bibliography{refs_kevin}

%

\end{document}